\documentclass{ws-jai}
\usepackage[flushleft]{threeparttable}
\usepackage[dvipsnames]{xcolor}
\begin{document}

\catchline{}{}{}{}{} 

\markboth{Ningxiao Zhang}{Metrology for Measuring Custom Periodicities on Diffraction Gratings}

\title{Metrology for Measuring Custom Periodicities on Diffraction Gratings}

\author{Ningxiao Zhang$^{1}$, Randall McEntaffer$^{1}$, Ross McCurdy$^{1}$ and Casey DeRoo$^{2}$}

\address{
$^{1}$Department of Astronomy and Astrophysics, The Pennsylvania State University, University Park, PA 16802, USA\\
$^{2}$Department of Physics and Astronomy,  The University of Iowa, Iowa City, Iowa 52242, USA}

\maketitle


\begin{history}
\received{(to be inserted by publisher)};
\revised{(to be inserted by publisher)};
\accepted{(to be inserted by publisher)};
\end{history}

\begin{abstract}
We present a new, inexpensive, bench-top method for measuring groove period over large areas with high mapping resolution and high measurement accuracy, dubbed the grating mapper for accurate period (GMAP). The GMAP has the ability to measure large groove period changes and non-parallel grooves, both of which cannot be measured via optical interferometry. In this paper, we detail the calibration and setup of the GMAP, and employ the instrument to measure three distinct gratings. Two of these measured gratings have customized groove patterns that prevent them from being measured via other traditional methods, such as optical interferometry. Our implementation of this tool achieves a spatial resolution of 0.1 mm$\times$0.1 mm and a period error of 1.7 nm for a 3 $\mu$m size groove period.
\end{abstract}

\keywords{diffraction grating; customized groove patter; period; mapping; metrology.}

\section{Introduction}
\noindent Spectroscopy is a fundamental exploratory tool in the study of astronomy to investigate plasma components, density, and temperature in stars, planets, interstellar material, etc. A diffraction grating is an important spectroscope component exhibiting a periodic structure that diffracts light into constituent wavelengths. Transmission gratings and reflection gratings are two types of diffraction gratings. The reflection grating normally shows strength in higher diffraction efficiency and a wider range of spectral coverage. Usually, the reflection grating is patterned with parallel grooves with a constant period. However, to achieve a higher spectral resolving power, our group is working on a non-parallel, radial grooves grating fabrication method \citep{randall2013,drew2018,randall2019,jake2020}. This novel grating requires a customized groove pattern with a variation of the period of as little as a few nanometers over centimeters of groove length. To qualify the fabrication method of this special reflection grating, we require an accurate mapping method for a customized groove pattern on this spatial scale.
 
Accurate groove period measurement, practical scanning time, and ability for scanning non-parallel grooves are three requirements for mapping the period distribution of a grating with customized groove pattern. Scanning electron microscopy (SEM), scanning probe microscopy (SPM), and optical interferometry are three traditional methods for measuring groove period of this type of gratings with parallel grooves \citep{tatsuo1987, misumi2003, voronov2017}. However, these methods fall short of achieving the three requirements. SEM is a developed microscopy method used for several decades to directly image the surface topography of a sample. It has a large magnification level ranging from 100 to 100,000 \citep{hansen2006}. The resolution can be 2 nm for the highest magnification \citep{hansen2006}. However, the actual resolution will be much lower when scanning a big feature under a lower magnification. Typical tool accuracy is $\sim$5\% of a feature size. Thus, the SEM method has limited accuracy when the feature size is large and its small image area limits the scanning speed. SPM, including scanning tunneling microscopes and atomic force microscopes, can offer topography data on very fine surfaces. Basically, this method uses a sharp probe to scan over the sample with an area up to 100 $\mu$m $\times$ 100 $\mu$m \citep{hansen2006}. SPMs apply the piezoelectric effect to maintain a small distance between the probe and the sample. Thus, this technology can measure the depth variation on the surface of the sample and the vertical resolution can reach the sub-nanometer level \citep{hansen2006}. However, SPMs do not have the same accuracy in measuring the lateral size of a feature (typically ~2\%), which is the critical dimension for groove period measurements. SPM is usually used to scan even a smaller area than SEM, making scanning gratings with area of square centimeter impractical. Optical interferometry works very well for mapping constant groove density over large areas with high accuracy quickly \citep{hutley1982, palmer2002, voronov2017}. This method can quickly map a sample with a large area (such as a size of 100 mm $\times$ 100 mm), which is much beyond the ability of SEM and SPM. This method is based on the technology of measuring a phase change of two reflected light beams when the diffraction grating satisfies the Littrow condition \citep{samuel2017}. The variation of the phase can be transformed to the variation of the groove period of the grating. The resolution of this method can reach sub-nm \citep{hansen2006}; however, the period variation can be resolved over only a very narrow period range which is related to the wavelength of the laser. A custom grating with large period variation or non-parallel grooves cannot be adequately measured with this method. For example, a grating with a 3 $\mu$m groove period and a period variation of 20 nm will have too large an angular variation to be measured interferometrically. In addition, the horizontal scanning resolution will be significantly reduced when the Littrow angle is large, as required to measure gratings with fine periods ($<$1000 nm) with commercial interferometers ($\lambda$ = 632.8 nm, typically). 

Other studies show that a the laser reflection tool can demonstrate high accuracy in measuring grating period variations \citep{dewey1994, jungki2017}. \citet{dewey1994} designed a laser reflection (LR) period mapping tool to verify the High-Energy Transmission Grating for the {\it{Chandra X-ray Observatory}}. They measured the spot motions of a reflected beam and a diffracted beam to calculate the grating period variation. \citet{jungki2017} improved the LR tool by adding a normal-incidence beam to account for the surface height variations. Their system has a higher accuracy and better repeatability compared to the LR tool. However, their design requires a well-calibrated grating with a similar groove period as a reference. This design is impractical for measuring groove period of a grating with unknown period or customized groove patterns. Thus, a new method capable of measuring custom-period gratings is still desirable.

In this paper, we present a new, inexpensive, bench-top method for measuring groove period variations over large areas in a reasonable time, the grating mapper for accurate period (GMAP). The details of using the GMAP to map the groove period are described in Section 2. The GMAP accurately measures groove periods, has the ability to identify non-parallel grooves, and can be used to scan a large area. To achieve these qualities, the GMAP needs to be well-calibrated. A walkthrough of this process is explained in Section 3. We employed the GMAP to measure three gratings with different designed patterns. The test configuration of these three measurements is discussed in Section 4. Finally, a discussion and summary of the test results are provided in Section 5.

\section{Test Methodology}
\begin{figure}[htp!]
\centering
\includegraphics[width=0.6\textwidth]{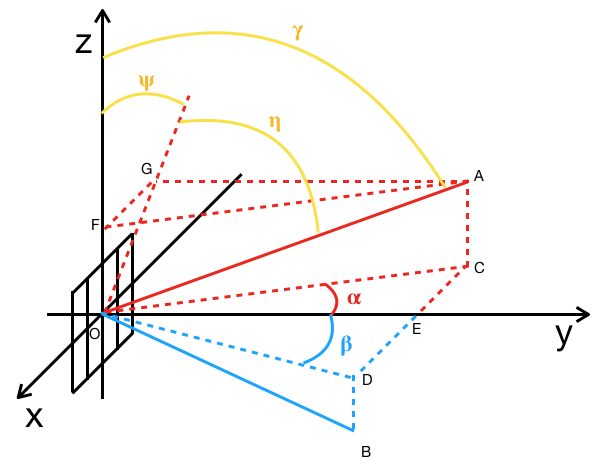}
\caption{Geometry of a reflection grating. A grating is set on xz-plane and the groove direction is the same as the z-axis. Light from point A is incident on the grating at point O. The incident beam is offset from the groove direction by an angle $\Psi$ (yaw) and strikes the grating at an angle $\eta$ (90$^o$ - angle of incidence). $\alpha$ and $\beta$ are azimuthal angles of incident light and diffracted light, respectively. $\gamma$ is the angle between the incident light and the groove direction.}
\label{fig1}
\end{figure}

\noindent The GMAP, similar to the LR tool, calculates the groove period by localizing the direction of diffracted orders, but the GMAP can directly measure the diffraction angle without using an extra grating as a reference. The general geometry of a reflection grating is shown in Figure \ref{fig1}, and the generalized grating equation is:
\begin{equation}
\sin \alpha + \sin \beta = \frac{n \lambda}{d \sin \gamma}\label{eq:1}
\end{equation}
where $\gamma$ is the angle between the incident beam (AO in Figure 1) and the groove direction (axis z in Figure 1), $d$ is the period, $\lambda$ is the wavelength of the incident beam, $\alpha$ is the polar incidence angle, and $\beta$ is the polar angle of the diffracted light (OB in Figure 1) for nth order \citep{cash1982, randall2013, frassetto2017}. In Figure 1, the pitch, yaw, and roll axes of the test grating are x, y, and z, respectively. Based on this geometry, we can calculate the grating period ($d$) by measuring the angles ($\alpha$, $\beta$, and $\gamma$) and using a stable laser beam ($\lambda$). 

According to the generalized grating equation (\ref{eq:1}), we need to measure angles $\alpha$, $\beta$, and $\gamma$ to calculate the groove period $d$. To simplify the measurement, we set the grating in an in-plane mounting such that $\gamma =  90^o$ and $\alpha = 0^o$. Then, the grating equation is:
\begin{equation}
d = \frac{n \lambda}{ \sin \beta }. \label{eq:2}
\end{equation}
In this case, we only need to measure $\beta$ to calculate the groove period. In the GMAP, we use a rotation stage to measure this diffraction angle $\beta$ and a camera to record the spot of the diffracted beam at nth order. In the initial GMAP setup (left panel of Figure 2), the laser beam is set parallel to the grating normal while $\beta$ is the azimuthal angle of the diffracted light in nth order. The reflected light beam (0th order) returns back to the laser. Then we rotate the rotation stage $\theta$ degrees, such that the 0th order is placed in the same spot on the camera as nth order (right panel of Figure 2). In this case, we have $\beta= 2 \theta$ as the angle of incidence and reflection are the same. Placing $\beta$ back into equation (\ref{eq:2}), we obtain the groove period on the grating where the laser beam is incident. 

\begin{figure}[htp!]
\centering
\includegraphics[width=0.9\textwidth]{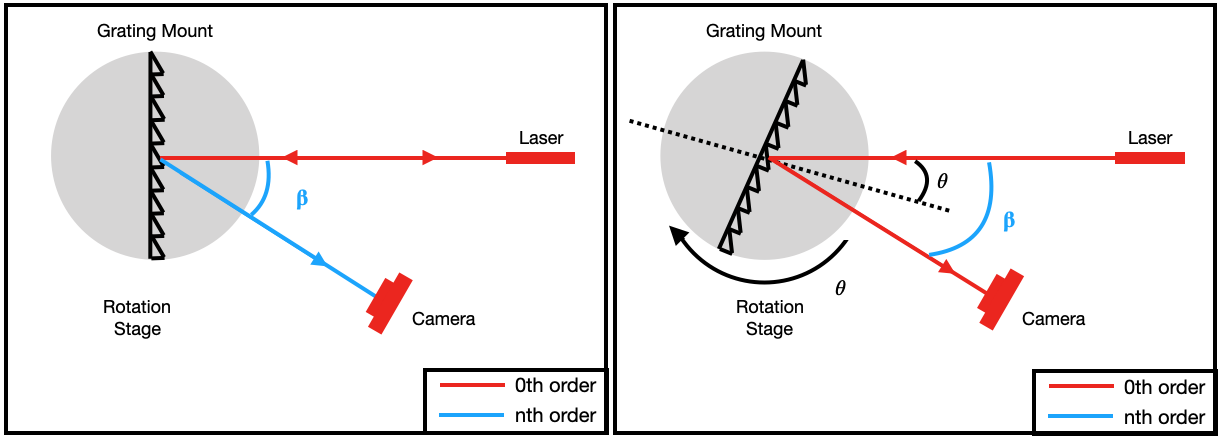}
\caption{Left - The geometry of the setup of the GMAP, where the camera is set at the nth order diffracted light beam and the 0th order light beam is reflected back to the laser. Right - After rotating the rotation stage with a degree of $\theta$, 0th order lands at the same camera pixel as nth order in the left panel. The dashed line is the grating normal.}
\label{fig2}
\end{figure}

With this process, we can make an absolute measurement of the period at any spot on the grating. However, in our implementation of the GMAP, it takes approximately 20 seconds to repeat this process. At this speed, measuring a 20 mm $\times$ 20 mm grating at 0.1 mm resolution would require roughly 9 days, which is impractical. In practice, we conduct this measurement once, and employ this spot as a reference to deduce changes in the groove period relative to this reference point. According to equation (\ref{eq:2}), a change in the groove period will change the angle of the diffracted light beam ($\beta$). Thus, we only need to measure how much the azimuthal angles of the diffracted light in nth order change because the distance of the grating and the camera is fixed in this process. The changing angle can be calculated by measuring the number of pixels the diffraction spot shifts. This is a fast way to map the groove period of a test grating without rotating the stage, and the entire process only requires half the time with the same mapping resolution requirement. 

This process assumes the grating surface is flat; however, in actuality, the grating surface is not ideally flat. The surface figure of the grating will cause additional shifts in the diffracted light beam in both the horizontal and vertical direction. The direction of the diffracted light beam at -nth order and nth order has the same shift due to the surface curvature and opposite shift from the changing groove period. Thus, we set a second camera on the -nth order as shown in Figure 3. Subtracting the shift in degrees at nth order from the degrees at -nth order allows us to account for the effect of surface curvature. 

\begin{figure}[htp!]
\centering
\includegraphics[width=0.7\textwidth]{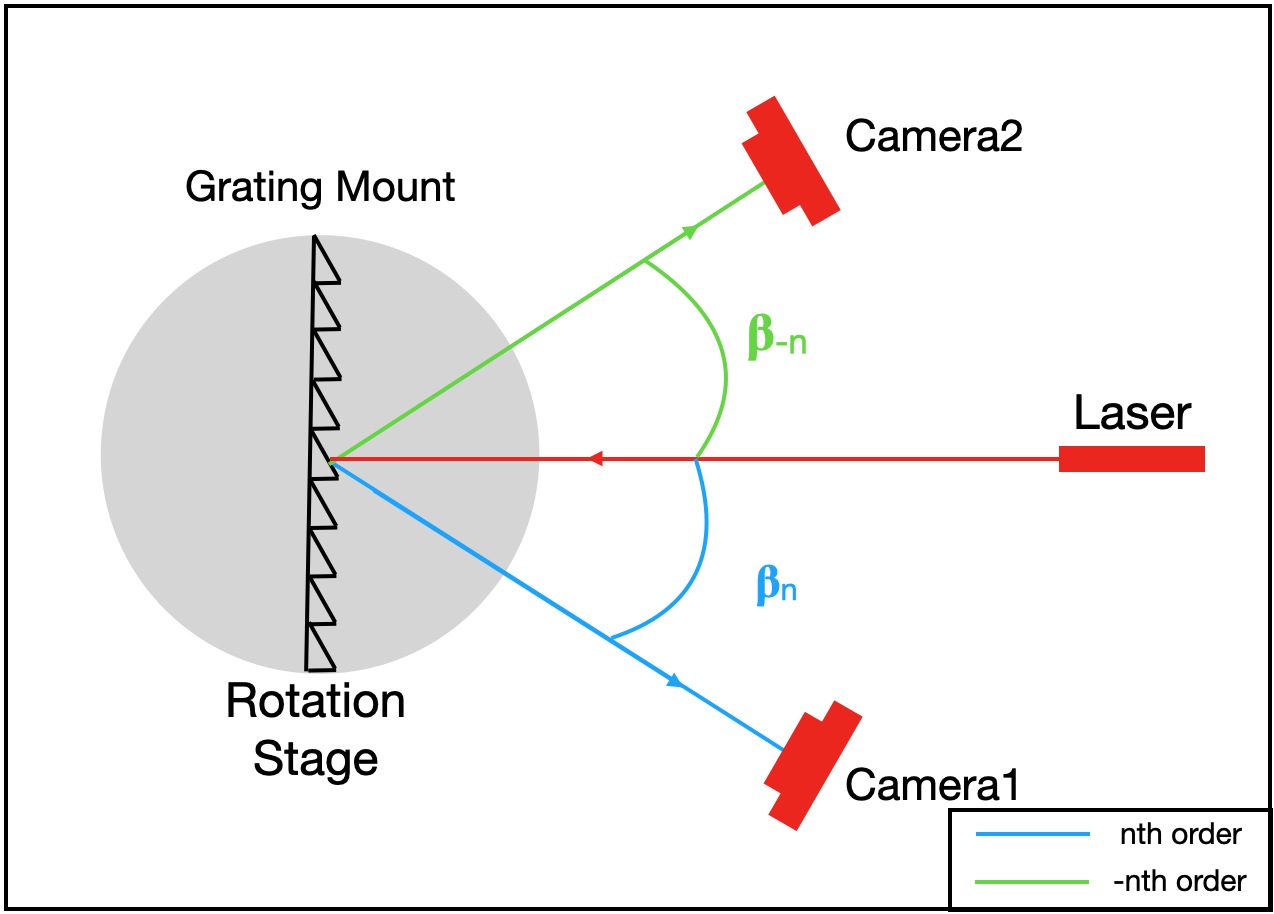}
\caption{The geometry of all components in the GMAP. The grating was set in an in-plane mounting on a rotation stage. Two cameras are set at nth and -nth orders and $\beta_{n}$ and $\beta_{-n}$ are the azimuthal angles of the diffracted light in the nth and -nth orders, respectively.}
\label{fig3}
\end{figure}

\section{The GMAP Setup}
The GMAP includes seven main components: a laser source, a rotation stage, two linear stages, a kinematic platform mount, and two cameras. The grating is held by a kinematic platform mounted on two linear stages, which are set in a horizontal/vertical configuration, for 2D scanning. And the linear stages are set on top of a rotation stage (left panel of Figure 4). Two cameras are set at nth and -nth diffraction orders (n is chosen depending on required accuracy.). An image of all components installed on a precision-tuned, damped optical table is presented in the right panel of Figure 4. 

\begin{figure}[htp!]
\centering
\includegraphics[width=0.9\textwidth]{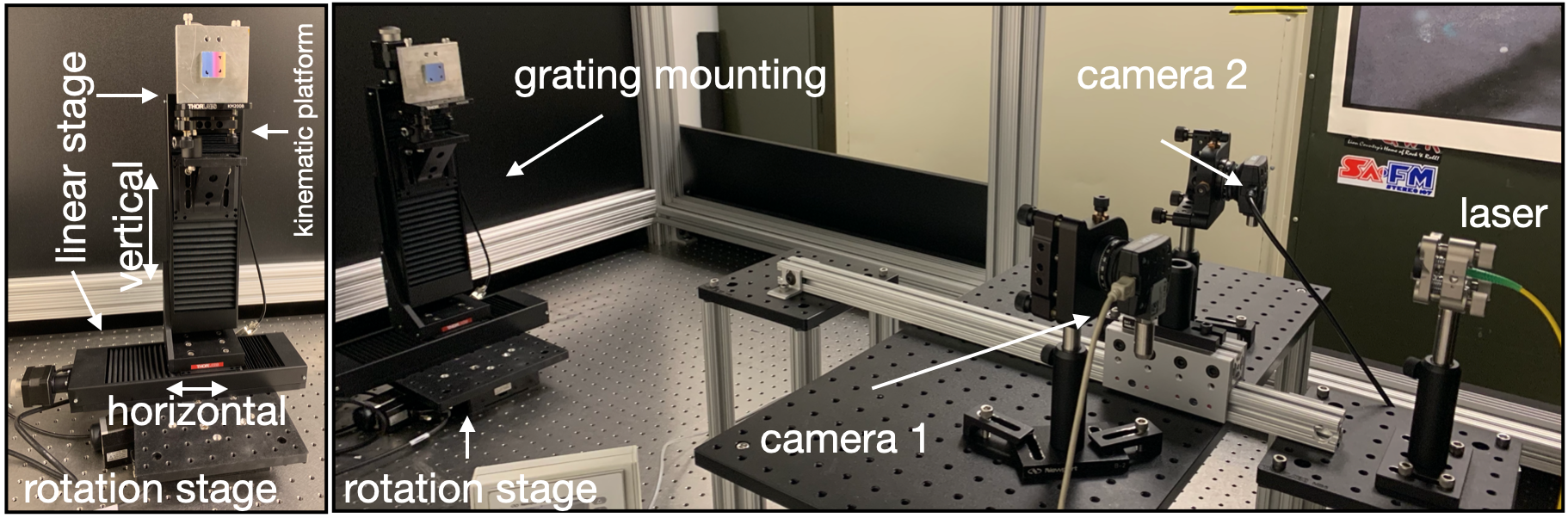}
\caption{Left - A grating mount includes a rotation stage and two linear stages. Right - An image of all the components of the GMAP.}
\label{fig4}
\end{figure}

The GMAP is set on a vibration dampened optical table from Newport to reduce the structural vibrations of the tabletop. We set the rotation stage as the reference of the GMAP. The motorized rotation stage is a URS100BPP model from Newport. The stage offers a 360$^o$ angular range, $\pm$4.4 mdeg bi-directional repeatability, and $\pm$8 mdeg accuracy. The machining accuracy of the alignment between the rotation axis and the center axis of the stage (eccentricity) is 0.4 $\mu$m. After placement of the rotation stage, there are three steps: installing the laser, initializing the grating, and setting up the cameras. Rotational degrees of freedom for these three components are shown in Figure 5. The rest of this section will describe the setup sequence and the uncertainty estimation at each step. 

\begin{figure}[htp!]
\centering
\includegraphics[width=0.8\textwidth]{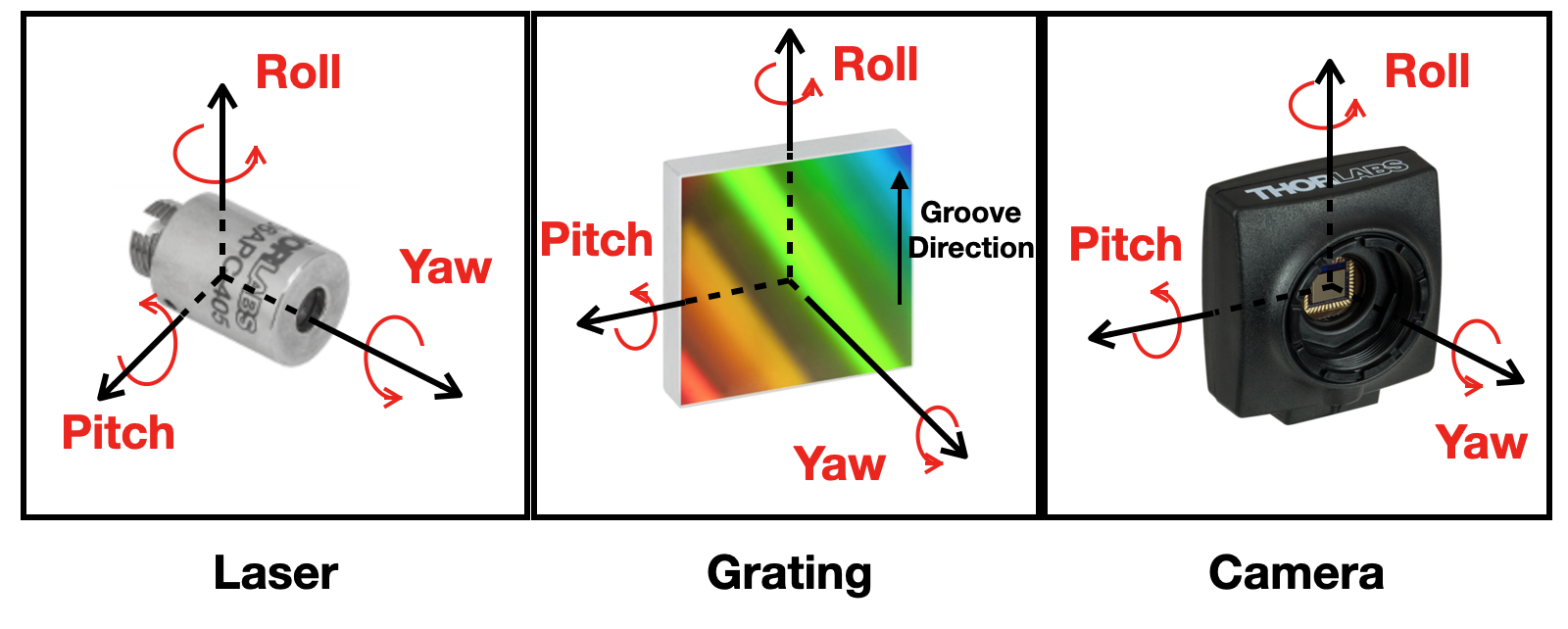}
\caption{Rotational degrees of freedom for the laser, the grating and the camera in the left, middle, and right panel, respectively. }
\label{fig5}
\end{figure}

\subsection{Installing the laser}

The laser system consists of a single-mode fiber-coupled laser, a single-mode patch cable with FC/PC connectors, a collimator, and a kinematic mount. The laser source has a center wavelength of $\lambda$ = 402.33$\pm$0.06 nm, where this uncertainty of the laser source is derived from a Gaussian fit to the laser spectrum (left panel of Figure 6). The resolution of the spectral data (scanned from a laser test report offered by Thorlabs) is not adequate for unambiguous determination of the center wavelength. Thus, we used a 3 $\mu$m period grating fabricated through electron-beam lithography to identify the center wavelength. The laser source system requires 1 hr to warm up at 25 $^o$C to be stable at 6 mW. According to the theoretical beam diameter graph offered from Thorlabs, the full-angle divergence of 0.027$^o$ makes the beam diameter 1 mm, 1.5 mm, and 1.75 mm at 1 m, 2 m, and 3 m, respectively, from the collimator.

The collimator is mounted on a kinematic mount, which is designed to provide high-resolution adjustment for roll and pitch. To ensure that the laser hits the same spot on the grating while rotating the stage, we need to set the laser light beam coincident with the rotation axis. To roughly align the laser source with the rotation axis of the rotation stage, we first set the laser beam to shoot horizontally along the grid of mounting holes on the table 1 m from the rotation axis (Figure 7). A camera was set at two positions, which are 0.4 m apart, on the optical table for fine adjustment. We adjust the roll and pitch of the laser beam through the kinematic mount such that the beam spot hits the same spot on the camera at these two distances. The camera will be removed after this step.

The right panel of Figure 6 shows the laser spot on a Thorlabs CMOS camera, which has 1280 $\times$ 1024 pixels and each pixel is a 5.2 $\mu$m $\times$ 5.2 $\mu$m square. The center coordinate of the spot is identified as the centroid of a Gaussian fit to the x and y intensity distributions as recorded on the camera. To estimate the uncertainty of this localization, we performed a stability test, recording the centroid of the laser in a fixed position 1000 times. This test yields a repeatability of $\pm$ 1 pixel, which we interpret as the uncertainty in our centroid measurement. This uncertainty is equivalent to an uncertainty of 0.001$^o$ for the roll and pitch of the laser. And the position tolerance of the optical table hole is 0.048''$\pm$0.017'', which has a negligible effect on the alignment. This indicates the laser beam has an error of offset to the rotation axis of $\sim$20 $\mu$m at a 1 m distance. During operation of the GMAP, rotations are limited to less than 5 degrees, which would cause a maximum spot shift on the grating of 0.1 $\mu$m. We assume that the groove period can be approximated as constant in this small area, and hence this beam displacement has a negligible effect on determining the reference period of the grating. 

\begin{figure}[htp!]
\centering
\includegraphics[width=0.8\textwidth]{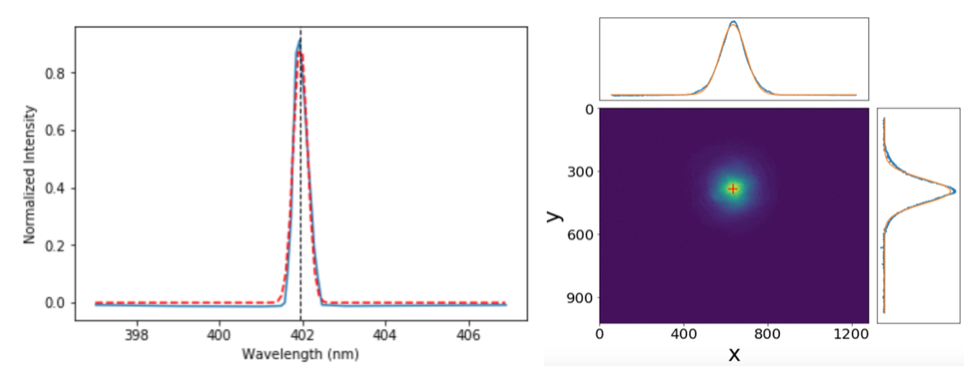}
\caption{Left - The spectrum of the laser source. The red dashed line is a Gaussian fit with a standard deviation of 0.06 nm. Right - An image of a laser spot on the camera with a 1 m distance from the collimator. The x and y units are camera pixels. The top and right panels show the integrated intensity of each column and row fit by Gaussians. The spot center is marked by the red cross where the coordinates are the peaks of the two fitted Gaussians.}
\label{fig6}
\end{figure}

\begin{figure}[htp!]
\centering
\includegraphics[width=0.6\textwidth]{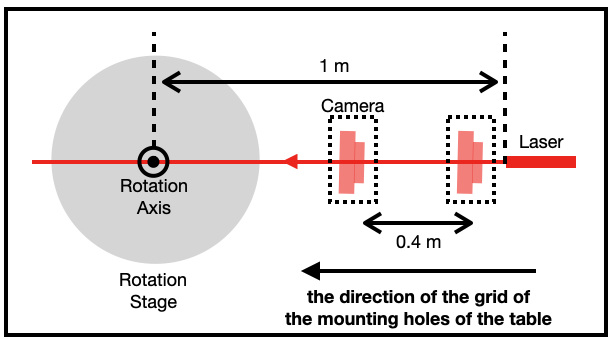}
\caption{The geometry of aligning the laser source with the rotation stage. The rotation axis is set 1 m from the collimator. The camera is set at two distances, which are 0.4 m apart, along the direction of the grid of the mounting holes of the table.}
\label{fig7}
\end{figure}

\subsection{Initializing the grating}
The grating mount includes two linear stages and a kinematic platform mount.  To achieve 2D scanning, the two linear stages are set in a horizontal/vertical configuration (left panel of Figure 4) such that the grating normal is parallel to the laser beam. Each of them has a travel range of 150 mm and a bi-directional repeatability of 1 $\mu$m. The chosen step size of the linear stages depends on the laser beam size on the sample. A small beam size and a small step size can offer high spatial resolution. However, there is a trade-off between spatial resolution and scanning time. Our spatial resolution is set based on the limitation of our laser beam size of 1 mm. In this case, a 0.1 mm step size ensures that two continuous measurements have less than 90\% same area coverage. This 90\% threshold can be adjusted based on a scanning time requirement. The grating is mounted to the linear stage which is then mounted to a kinematic platform through a slot hole. The slot hole provides $\sim$25 mm of adjustment in the horizontal direction while the kinematic platform mount provides $\pm3^o$ tip/tilt adjustment for correcting yaw and pitch (left panel of Figure 4). To make sure the laser is incident on the same spot on the grating while rotating, we need to set the grating so that the central groove is coincident with the axis of the rotation stage. In the previous step, we already set the laser through the axis of the rotation stage; therefore, setting the laser beam coincident with the surface of the grating is equal to setting the axis of the rotation stage on the grating surface (left panel of Figure 8). We first set the grating surface roughly on the path of the laser beam such that half of the beam passes through directly to the camera and the other half is reflected by a small degree. We use a camera to record the location of these two spots and we adjust the grating surface until these two spots overlap each other. This indicates that the grating surface is parallel to the laser beam within 0.03$^o$ (assuming 0.5 mm half beam size at 1 m distance) and coincident with the axis of the rotation stage. Then we adjust pitch and yaw of the grating (the roll correction of the grating will be discussed during the camera setup). 

To adjust the pitch, we rotate the grating 90$^o$ such that the laser beam is parallel to the grating normal. Slight misalignments of pitch will cause the reflected laser spot to be offset from the collimator aperture (4.5 mm) in the vertical direction. We adjust the kinematic mount such that the laser beam reflects back into the aperture of the collimator. We assume that we can distinguish the offset of the laser beam by half of the beam size (assuming 1.5 mm beam size at 2 m distance), which gives a pitch uncertainty of 0.1$^o$. 

Next, we need to calibrate the yaw of the grating. We ultimately need to place the grating at $\alpha = 3^o$ in our setup to ensure that the camera does not block the laser (the geometry of this initial position is shown in the right panel of Figure 8). To correct the yaw of the grating, we temporarily set a camera (details are in the camera setup section) 1 m from the grating at 0th order and record its X and Y pixel coordinates (left panel of Figure 9). Then we rotate the rotation stage such that the 1st order diffraction spot hits the camera with the same X pixel as we recorded for 0th order (right panel of Figure 9). Any residual Y pixel offset between the diffracted order and 0th order is due to yaw misalignment. We therefore adjust the kinematic platform mount until the 1st order spot has the same coordinates as the 0th order. Due to the uncertainty of the center of the spot of 1 pixel, the uncertainty of the yaw of the grating is 0.0003$^o$. 

Setting the pitch and yaw of the grating in this way will ensure that the angle $\gamma$ is as close to $90^o$ as possible. Based on this setup, we calculate the uncertainty of the angle $\gamma$ from the equation:  
\begin{equation}
\cos \gamma = \cos \Psi \cos \eta \label{eq:3}
\end{equation}
where the $\Psi$ is the yaw of the grating and $\eta$ is the pitch of the grating (shown in Figure 1). According to the uncertainties calculated above, we have $\Psi = 0 \pm 0.0003^o$ and $\eta = 90 \pm 0.1^o$. This results in $\gamma = 90 \pm 0.1^o$ where the error is calculated from Monte Carlo simulation. 

\begin{figure}[htp!]
\centering
\includegraphics[width=0.9\textwidth]{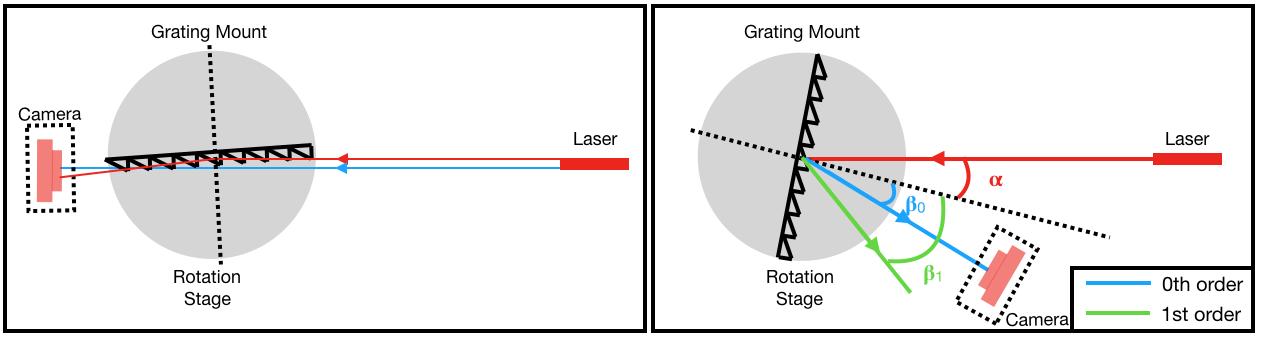}
\caption{Left - The geometry of aligning the grating with the laser beam. The surface of the grating is roughly parallel to the laser beam. The red line shows the light path of the reflected spot on the camera and the blue line represents the light path of the direct beam. Right - The initial geometry when we correct pitch of the grating. }
\label{fig8}
\end{figure}

\begin{figure}[htp!]
\centering
\includegraphics[width=0.6\textwidth]{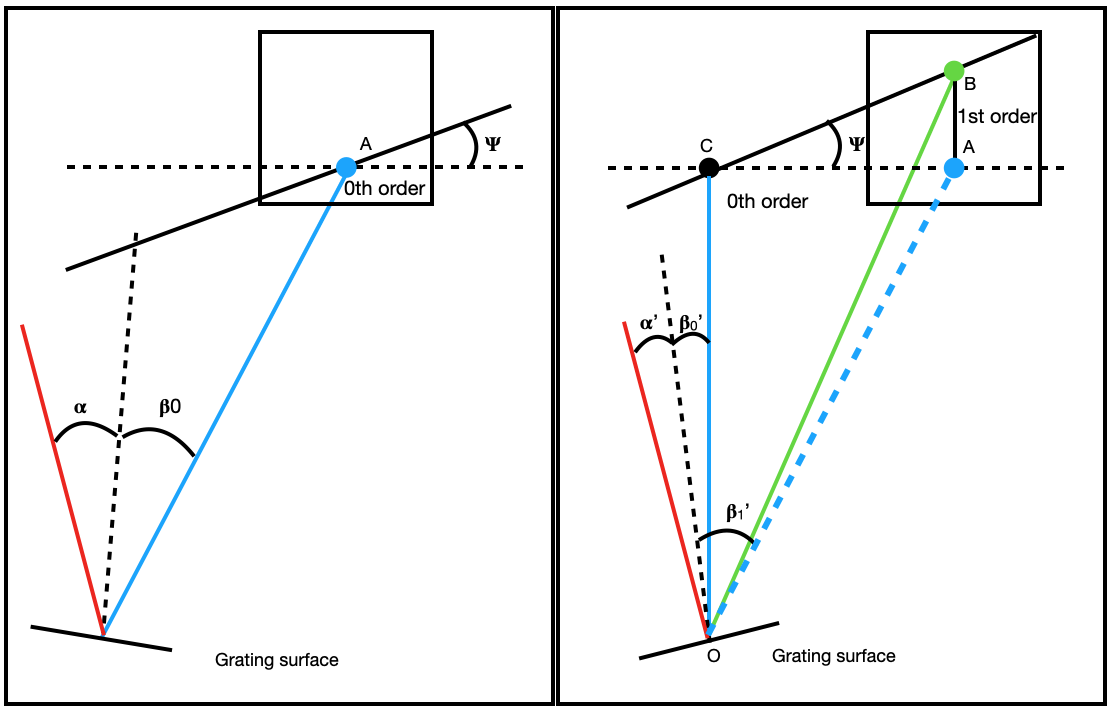}
\caption{Left - The initial geometry of setting grating yaw. The 0th order light beam is coincident on the camera. Right - After rotating the rotation stage, the 1st order light beam is incident on the camera with the same X pixel coordinate but slight misalignments of yaw will cause the different Y pixel coordinate.}
\label{fig9}
\end{figure}

\subsection{Camera setup}
The diffraction spots are recorded using two of the CMOS cameras described in Section 3.1. Each camera is fixed on a high precision 6-axis kinematic mount for the adjustment of roll, yaw, and pitch. The kinematic mount is used to align the camera normal to the direction of the incident beam. Before installing the camera, we use a mirror to adjust the roll and pitch of the kinematic mount (left panel of Figure 10), such that the beam reflected by the mirror returns to the same spot as the incident laser beam on the grating. We assume that our eye can easily distinguish two separated spots by a distance of the beam size. Thus, the uncertainty of the pitch and roll of the mirror are 0.1$^o$ with a 1.75 mm beam size given a 2 m optical path of the laser beam. Motivated by the repeatablility and stability specified by ThorLabs, we assume that the mount holds the camera plane within one degree of the aligned mirror plane, which will cause a maximum shift on the camera of 1.6 $\mu$m. This is a negligible effect (less than 1 pixel) given the 5.2 $\mu$m pixel size.

We adjust the roll and pitch of the mount such that the beam reflected by the mirror returns to the same spot as the incident laser beam on the grating. This correction makes sure the mirror normal is parallel to the diffracted laser beam. 

The mirror is then replaced by the camera (right panel of Figure 10). A camera yaw offset will cause the spot to drift in the X and Y directions. To adjust the yaw of the camera, we rotate the rotation stage such that the reflected spot moves horizontally across the camera (left panel of Figure 11). We adjust the yaw until the spot always has the same Y pixel value (right panel of Figure 11). Due to the uncertainty of the spot centroid (1 pixels), the uncertainty of the yaw of the camera is 0.07$^o$ given the 800 pixel distance over which the spot is measured. 

During the measurement, we also need to calibrate the number of pixels the spot moves on the camera as a function of the number of degrees of rotation on the grating stage. Thus, we record the pixel coordinate and the degree of the rotation stage for 10 spots after yaw correction. Based on a simple linear regression model, we can transfer the shift of the center pixel value to the actual degree shift of the spot on each camera. This method has the benefit of not needing to measure the distance between each camera plane to the grating surface.

\begin{figure}[htp!]
\centering
\includegraphics[width=0.9\textwidth]{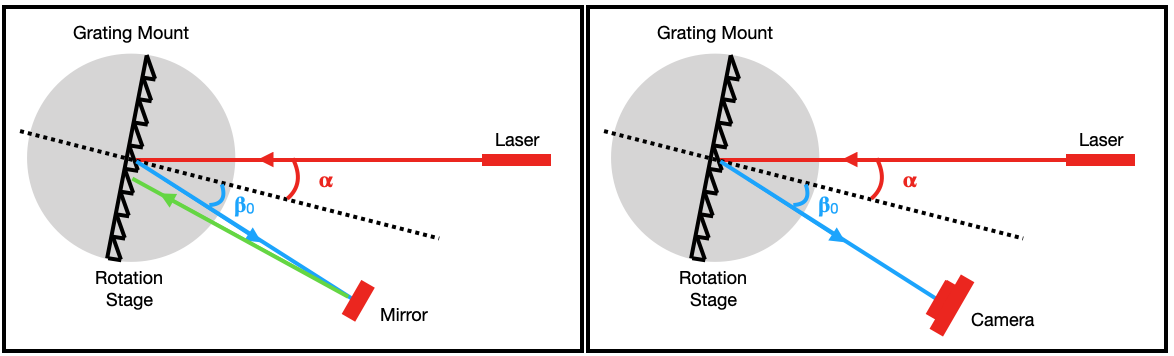}
\caption{Left - The geometry of setting roll and pitch of the kinematic mount which is used to hold the camera. The blue line represents the diffracted light beam and the green line shows the reflected light beam from the mirror. Right - After correcting the roll and pitch of the mount, the mirror is replaced by the camera.}
\label{fig10}
\end{figure}

\begin{figure}[htp!]
\centering
\includegraphics[width=0.9\textwidth]{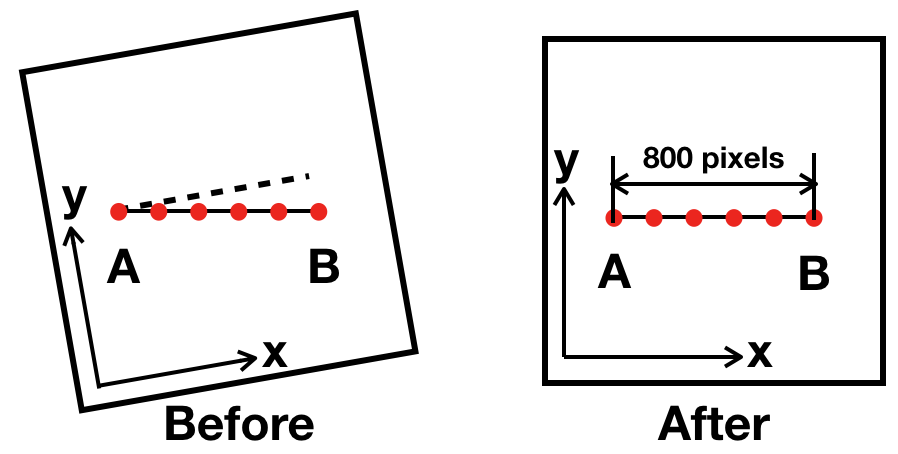}
\caption{The camera before and after yaw correction. The dashed line represents the x direction of the camera. The spots A and B are separated by a distance of 800 pixels in the x direction.}
\label{fig11}
\end{figure}

Using the above procedure, we place and calibrate the nth and -nth order cameras. We can then perform the final alignment and set the roll of the grating such that $\alpha = 0^o$. First we record the coordinates of the nth and -nth diffracted order on the two cameras. Then, we rotate the grating such that the 0th order spot hits the same coordinate as nth order and record this angle as $\theta_{+n}$. We then adjust roll in the opposite direction such that 0th order hits the same coordinate as -nth order and record this degree as $\theta_{-n}$. Finally, we rotate back $(\theta_{+n}+\theta_{-n})/2$ degrees to remove any roll offset. The uncertainty of the center of the spot is 1 pixel at 1 m distance, thus giving a grating roll uncertainty of 0.0002$^o$.

\subsection{System uncertainties}
Based on the accuracy of the equipment used, the uncertainty of angle $\theta$ is determined by the roll of the grating (0.0002$^o$), the uncertainty from the rotation stage (0.0044$^o$), and the uncertainty of the spot centroid (0.0003$^o$). This gives a $\theta$ ($\beta=2\theta$) uncertainty of 0.0044$^o$. Based on equation (\ref{eq:2}), this results in a groove period uncertainty of $\pm$ 3.4 nm/$n$ when measuring a 3 $\mu$m size groove period with our laser. All the uncertainties of the GMAP are shown in Table 1.

\begin{table}
\begin{center}
\begin{tabular}{ |p{5cm}|p{2.5cm}|  }

 \hline
 Error Terms & Uncertainties\\
 \hline
 Rotation Stage Repeatability & $\pm0.0044^o$ \\
 Rotation Stage Eccentricity & $\pm0.4\mu$m \\
 Laser Wavelength & $\pm0.06$nm \\
 Centroid of the Laser Spot & $\pm1$ pixel \\
 Laser Source Pitch & $\pm0.001^o$ \\
 Laser Source Roll & $\pm0.001^o$ \\
 Laser Beam Offset & $\pm20\mu$m \\
 Grating Pitch & $\pm0.1^o$ \\
 Grating Yaw & $\pm0.0003^o$ \\
 Grating Roll & $\pm0.0002^o$ \\
 Camera Pitch & $\pm0.1^o$ \\
 Camera Yaw & $\pm0.07^o$ \\
 Camera Roll & $\pm0.1^o$ \\
 $\alpha$ & $\pm0.0002^o$ \\
 $\beta$ &$\pm0.0088^o$ \\
 $\gamma$ & $\pm0.1^o$ \\
 \hline
\end{tabular}
\caption{A summary of the systematic uncertainties present in the GMAP set-up process. The resulting uncertainty in determining a 3 micron groove period is 3.4 nm/$n$, where $n$ is the working diffraction order of the system.}
\end{center}
\end{table}

\section{Measurements}
We have tested the efficacy of the GMAP with three gratings. These gratings, shown in Figure 12, are designed with different patterns to test different mapping abilities of the GMAP. First, we test it with a grating of parallel grooves and constant groove period (left panel of Figure 12). Second, we test it with a grating of parallel grooves and five different groove period sections (middle panel of Figure 12). Third, we test it with a grating of non-parallel, radial grooves (right panel of Figure 12). Cameras were set at +2nd and -2nd orders for each of these measurements.

The first grating (Grating 1, left panel of Figure 12) is a Thorlabs grating with designed constant groove period distance. The measurement of Grating 1 is designed to demonstrate that the result from the GMAP is consistent with the traditional optical interferometry method \citep{hutley1982}. The optical interferometry method is designed for mapping a grating with parallel grooves and constant groove period which is the same as the design of Grating 1. This method requires an assumption of the average groove period since an interferometer cannot measure absolute groove period accurately. Thus, we apply the mean groove period measured from GMAP as an average groove period for the optical interferometery method. Both mapping results of the GMAP and the optical interferometry method are shown in Figure 13. We mark Grating 1 with three marks on the top right, bottom left, and bottom right, respectively, to define the orientation. We extract groove period values from the red rectangle region of Grating 1 in each method. The histograms in Figure 14 show the groove period distributions from measurements of the GMAP and the optical interferometry method, respectively. The groove period is 3334.3$\pm$0.1 nm from the GMAP and 3334.3$\pm$0.9 nm from the optical interferometry method. Given a $\sim$1.7 nm system uncertainty, this result is consistent with the grating designed period of 3333.3 nm. Both methods show a constant groove period over most of Grating 1 and similar patterns on the left and right edges of the grating. However, the result from the optical interferometry method has more spatial detail (right panel of Figure 13) and larger groove period variation (right panel of Figure 14). Those details are smoothed out in the GMAP's result (left panel of Figure 13 and Figure 14), because the GMAP applies a $\sim$1 mm laser beam and each measurement is an average of the groove period in this 1 mm circle.

\begin{figure}[htp!]
\centering
\includegraphics[width=0.9\textwidth]{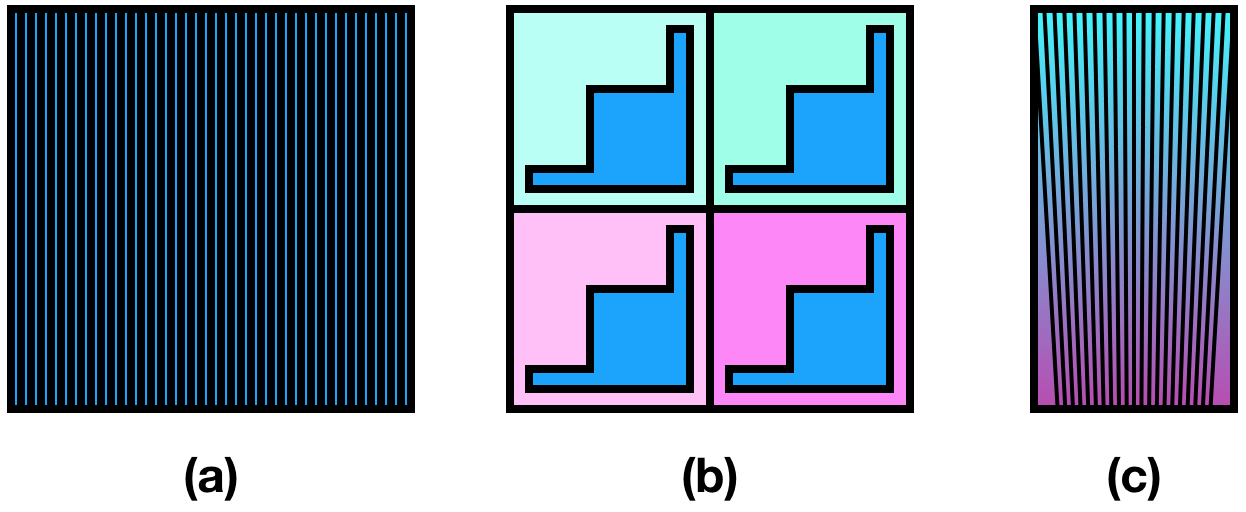}
\caption{Left - A Thorlabs grating with constant groove period distance of 3333 nm (Grating 1). Mid - The customized grating array with 4 square sections of groove period of 2990 nm, 2980 nm, 2960 nm, and 2970 nm. On each of these sections, there is a ``L" shaped pattern with constant groove period distance of 3000 nm (Grating 2). Right - The customized grating with converging grooves. The period distance smoothly changing from 3000 nm to 2950 nm (Grating 3).}
\label{fig12}
\end{figure}

\begin{figure}[htp!]
\centering
\includegraphics[width=0.9\textwidth]{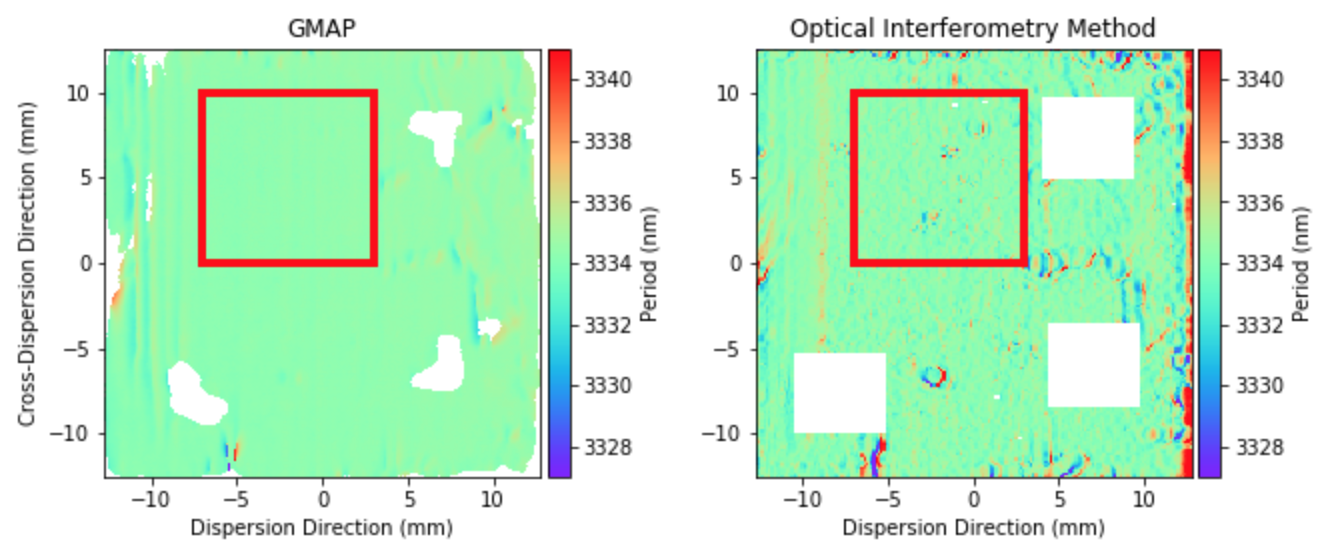}
\caption{Left - Mapping result of a Thorlabs grating with constant groove period measured by GMAP. Right - Mapping result of the same grating measured by optical interferometry method. The red rectangle in both images are the regions we extract the groove period values of the histograms in Figure 14.}
\label{fig13}
\end{figure}

\begin{figure}[htp!]
\centering
\includegraphics[width=0.9\textwidth]{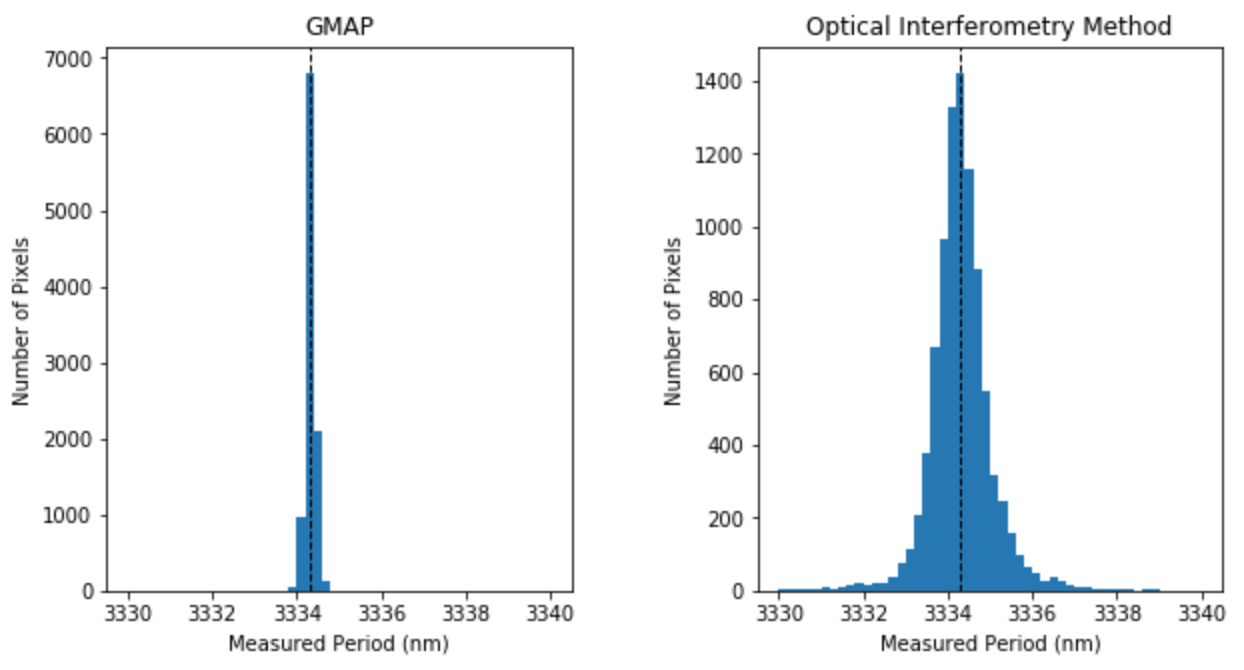}
\caption{Left - A histogram of the groove period value of the Grating 1 measured by GMAP. Right - A histogram of the groove period value of the Grating 1 measured by optical interferometry method. The dashed line in both histograms represent the assumed average groove period of 3334.3 nm.}
\label{fig14}
\end{figure}

The second grating (Grating 2, middle panel of Figure 12) is a custom-period grating with five different groove period sections. The four 10 mm $\times$ 10 mm squares have periods of 2990 nm, 2980 nm, 2960 nm, and 2970 nm, respectively. On each square, there is a ``L" shaped pattern with 3000 nm groove period. Based on our setup and the limitation of the camera size, the measurement of Grating 2 is designed to demonstrate that the GMAP can identify a minimum 10 nm and maximum 40 nm groove period change (40nm groove period change is twice beyond the limitation of the optical interferometry method) and distinguish features with size of 1 mm, 3 mm, and 5 mm in our setup. The mapping result of the GMAP is shown in the top panel of Figure 15. The straight edges between sections indicate that the image resolution is as small as the smallest feature size, 1 x 1 mm$^2$ per pixel. Moreover, we also extract groove period values from the rectangular regions in each section of Grating 2 as depicted by the black rectangles in the top panel of Figure 15. The histograms in the bottom panel of Figure 15 show the groove period distributions from measurements within these five black rectangles. The groove periods are 2989.9 $\pm$ 0.2 nm, 2980.2 $\pm$ 0.2 nm, 2969.8 $\pm$ 0.2 nm, 2960.2 $\pm$  0.1 nm, and 3000.0 $\pm$ 0.2 nm in the red, yellow, blue, green and purple regions from Figure 15, respectively. This result is consistent with the designed periods of 2990 nm, 2980 nm, 2970 nm, 2960 nm, and 3000 nm given a 1.7 nm system uncertainty. In addition to the accurate mapping result in each region, this test also indicates that the GMAP can identify groove period change in 10 nm increments and distinguish feature size as small as 1 mm. 

\begin{figure}[htp!]
\centering
\includegraphics[width=0.9\textwidth]{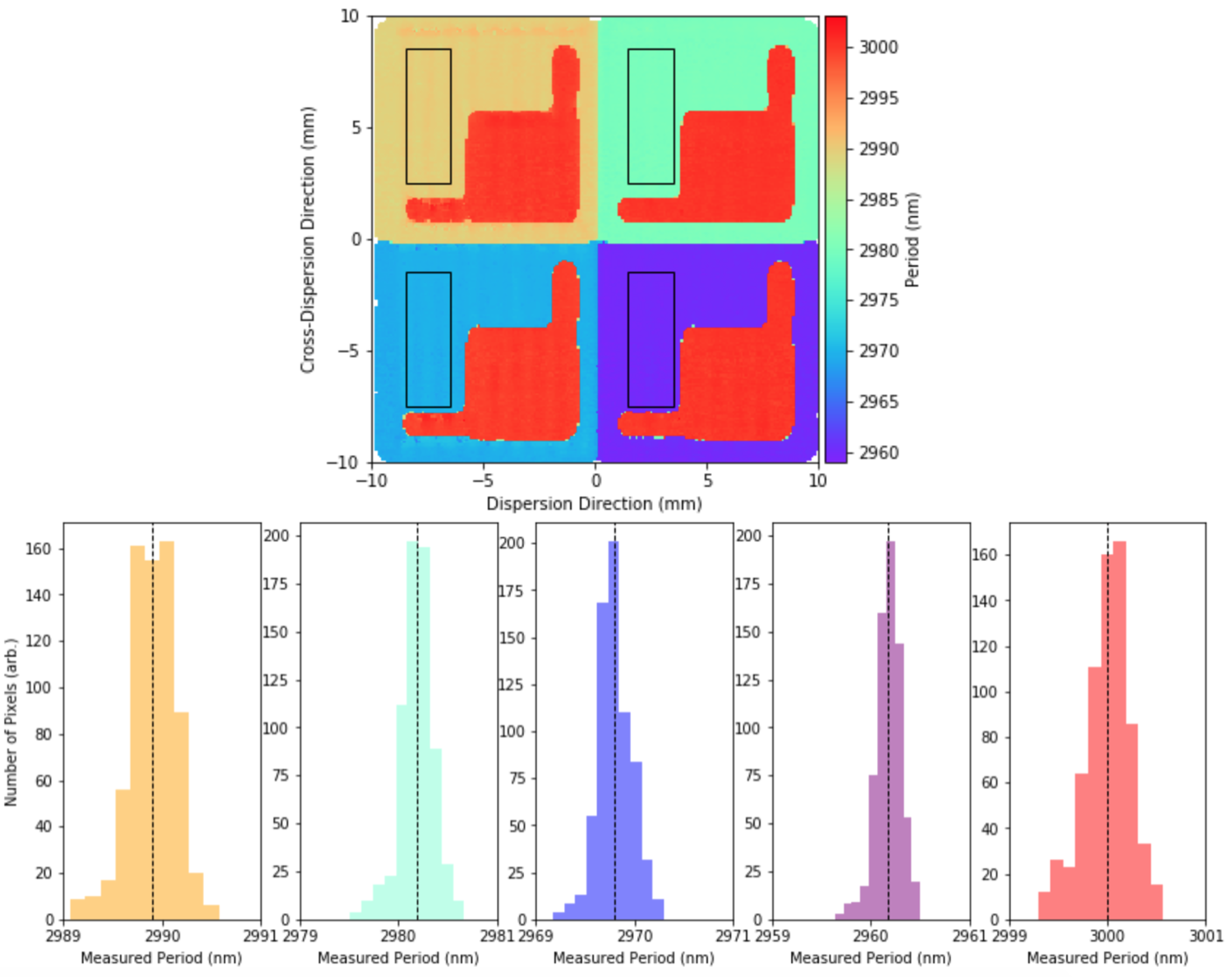}
\caption{Top - Mapping result of a customized grating with 4 square sections of different groove period distance measured by the GMAP. Bottom - Histograms of measured period distance in each of the black rectangular regions. The dashed lines represent the mean measured groove periods of 2989.9 nm, 2980.2 nm, 2969.8 nm, 2960.2 nm, and 3000.00 nm in the red, yellow, blue, green and purple regions, respectively.
}
\label{fig15}
\end{figure}

The third grating (Grating 3, right panel of Figure 12) is a custom-period grating with variable line spacing along the groove dimension, which results in a continuously variable groove period. This grating is 10 mm wide and 20 mm in the groove direction. Its groove period continuously changes from 3000 nm to 2950 nm. The measurement of Grating 3 is designed to prove that the GMAP can map non-parallel grooves gratings. The mapping result of the GMAP is shown in the right panel of Figure 16. The rainbow pattern on the grating shows the continously changing groove period from top to bottom. Furthermore, we average the groove period in every 10 rows (1 mm) of Grating 3 and plot them in the left panel of Figure 16. The X-axis represents the row distance to the center of the grating and the Y-axis represents the average groove period, while the vertical error bar on each data point represents the GMAP's systematic uncertainty of 1.7 nm in 2nd order. The red solid line displays the designed groove period of this grating with a slope of 2.50 nm/mm. The black dash line displays the fitted groove period with a slope of 2.48$\pm$0.01 nm/mm, which is within two standard deviations of the designed slope. This test indicates that the GMAP can measure continuous groove period change. 

\begin{figure}[htp!]
\centering
\includegraphics[width=0.9\textwidth]{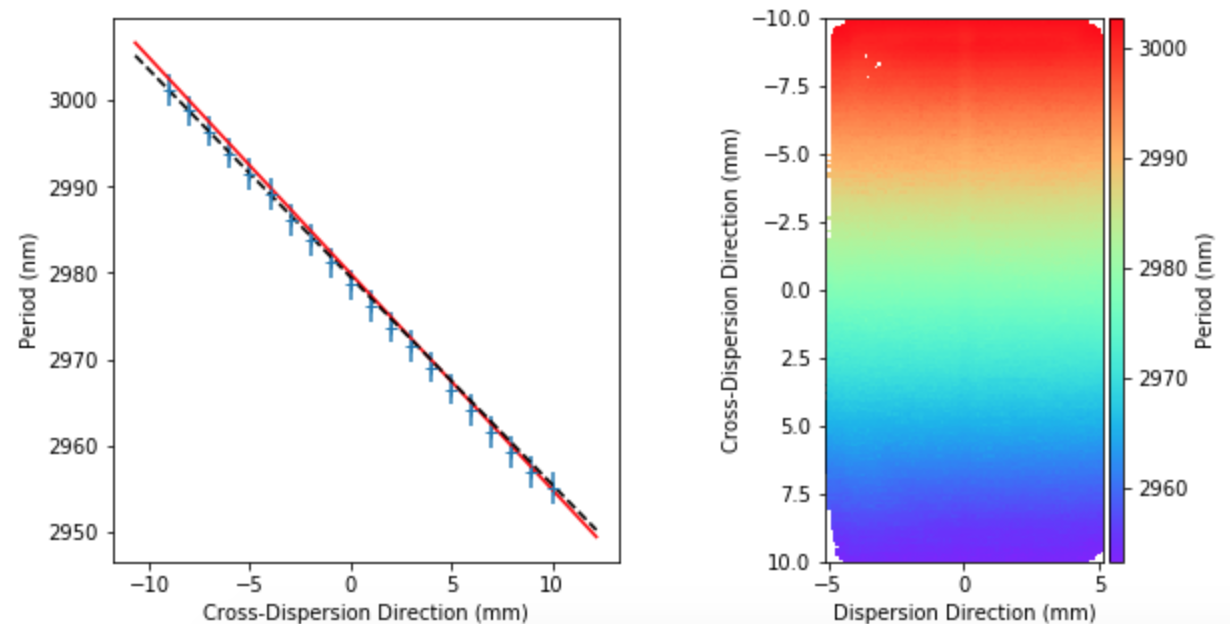}
\caption{Left - The measured period distance of the grating 3 in each row. The blue crosses represent the measured result. The red solid line shows the designed groove period and the black dash line represents the fitted groove period. Right - Mapping result of a customized grating with converging grooves measured by the GMAP.}
\label{fig16}
\end{figure}

\section{Discussion and Summary}

Our work shows that the GMAP is a versatile tool for measuring gratings and agrees with standard techniques of mapping groove period errors over a grating surface. As a proof of the methodology, we tested the GMAP on three gratings with different groove periods, different patterns, and different groove orientations. The summary of these results are shown in the Table 2. Compared to previous laser reflection tools, the GMAP offers the ability to measure the absolute groove period, simplifies the experiment setup and calibrations and improves uncertainty estimation. The GMAP also works well for gratings with large groove period changes and non-parallel grooves, for which the optical interferometry method does not work.

\begin{table}
\begin{center}
\caption{Summary of GMAP measurements on three gratings.}
\begin{tabular}{ |p{7cm}||p{4cm}|p{4cm}|  }

 \hline
 Tests & Design & Measured value\\
 \hline
 Grating 1 & 3333.3 nm & 3334.3 $\pm$1.7nm \\
 Red section on Grating 2 & 3000 nm & 3000.0$\pm$1.7 nm \\
 Yellow section on Grating 2 & 2990 nm & 2989.9$\pm$1.7 nm \\
 Green section on Grating 2 & 2980 nm & 2980.2$\pm$1.7 nm \\
 Blue section on Grating 2 & 2970 nm & 2969.8$\pm$1.7 nm \\
 Purple section on Grating 2 & 2960 nm & 2960.2$\pm$1.7 nm \\ 
 Groove period variation on Grating 3 & 2.50 nm/mm & 2.48$\pm$0.01 nm/mm \\
 \hline
\end{tabular}
\end{center}
\end{table}

The GMAP can be adapted for other applications. First, the GMAP could be used to map groove period distribution of ultraviolet or even x-ray gratings, which have much smaller groove periods. To achieve the ability to measure small groove periods (of order 200 nm), we can replace the optical laser source with an ultraviolet laser source. Second, the GMAP can be further modified by amplifying the angle of diffraction by setting the cameras in positions corresponding to higher orders. This will enhance the ability of the GMAP to distinguish small groove period changes. For example, GMAP gives an uncertainty of 3.4 nm/$n$ for a 3 $\mu$m grating at order $n$. Third, the GMAP can cover a wide range of groove period changes. This is useful for measuring aberration-correcting gratings or gratings enabling unique optical designs for spectrometers. In this paper, our setup has a limitation of 40 nm groove period change due to the limitation of the camera size. Switching to a bigger camera will improve the ability of the GMAP to cover a wider groove period change. Also moving the camera closer to the grating can improve this ability.  However, in this case, there is a trade-off between the range of the coverage and the accuracy of the measurement. 

\section{Acknowledgments}
This work was supported by NASA grants 80NSSC19K0661. We would also like to acknowledge internal funding from The Pennsylvania State University.

\bibliographystyle{ws-jai}
\bibliography{sample}
\end{document}